\documentstyle[aps,prl,floats,epsfig]{revtex}

\begin{document}
\draft

\twocolumn[\hsize\textwidth\columnwidth\hsize\csname
@twocolumnfalse\endcsname

\title{Field dependence of the electronic phase separation in Pr$_{0.67}$Ca$_{0.33}$MnO$_3$ by small
angle magnetic neutron scattering}
\author{Silvana Mercone, Vincent Hardy, Christine Martin, Charles Simon,}
\address{Laboratoire CRISMAT, UMR 6508 du CNRS et de l'Institut
Sup\'erieur de la Mati\`ere et du Rayonnement, 14050 CAEN, France.}
\author{Damien Saurel, Annie Br\^{u}let, }
\address{Laboratoire L\'{e}on Brillouin, CE SACLAY, 91191 Gif/Yvette, France}

\date{\today} \maketitle

\begin{abstract}
We have studied by small angle neutron scattering the evolution induced by
the application of magnetic field of the coexistence of ferromagnetism (F)
and antiferromagnetism (AF) in a crystal of Pr$_{0.67}$Ca$_{0.33}$MnO$_3$.
The results are compared to magnetic measurements which provide the
evolution of the ferromagnetic fraction. These results show that the growth
of the ferromagnetic phase corresponds to an increase of the thickness of
the ferromagnetic ''cabbage'' sheets.
\end{abstract}
]

\section{Introduction}

It was recently proposed that the ground state of manganites which displays
colossal magnetoresistive (CMR) properties \cite{review} could be an
electronic phase separation \cite{moreo,khomskii}. This is a very elegant
manner to interpret the CMR properties by percolation of a metallic
ferromagnetic phase in an insulating antiferromagnetic matrix. A small
change of the fraction or of the arrangement of the domains can induce
percolation. In order to go further in the percolation models, it is very
important to determine the size and the shape of the domains. Small angle
neutron scattering (SANS) is indeed a very powerful technique to study the
phase separation between a ferromagnet and an antiferromagnet since the
contrast between them is very large (the AF does not scatter at small angle)%
\cite{deteresa,radaelli,nous}. Among the manganites, the Pr$_{1-x}$Ca$_x$MnO$%
_3$ series is one of prime interest, because Pr and Ca are about the same
size and hence minimize the cationic size mismatch effect. Pr$_{1-x}$Ca$_x$%
MnO$_3$ magnetic phase diagram presents different states depending on the x
value. For the higher Mn$^{3+}$ contents (typically x=0.2), the compounds
are ferromagnetic at low temperature, and for larger x values (typically
x=0.4) they are orbital ordered, antiferromagnetic CE like type. In between,
the composition x=0.33, studied here, shows a mixing of F and AF phases as
shown by neutron diffraction\cite{jirak}. For these compositions close to
x=0.33, existence of the phase separation is now well proved by
magnetoresistance, magnetization and specific heat studies\cite{hardy,deac}.
Small angle neutron scattering has shown that at low temperature (below
30K), the structure is that of a ''cabbage'' with 2D sheets (stripes) of
about 2.5nm of thickness\cite{nous}. In the present paper, we propose to
study the influence of the application of a magnetic field on this phase
separation during percolation of the metallic phase.

\section{Experimental}

Using the floating-zone method with feeding rods of nominal compositions Pr$%
_{0.7}$Ca$_{0.3}$MnO$_3$, several-cm-long single crystals were grown in a
mirror furnace. Samples were cut out of the central part of these crystals
and were analyzed by EDS: their cationic compositions are homogeneous and
were found to be x= 0.33. Magnetization and transport measurements were
performed to check their quality. The samples were then powdered in order to
perform neutron diffraction. The powder diffraction patterns were recorded
in the G41 spectrometer in Orph\'{e}e reactor from 1.5K to 300K. It presents
at 1.5K both a F component and an AF one. Tc is about 100K and T$_N$ about
115K. The structures were refined using the Fullprof program, in the Pbnm
symmetry (a$_p\sqrt{2},$ a$_p\sqrt{2},$ 2a$_p$). The previous small angle
neutron scattering study\cite{nous} was performed on the same powdered
sample. In the present experiment, we have studied a part of the same sample
but in the form of a single crystal which was cut out of the rod without any
specific orientation.

Small angle neutron scattering were performed on PAXY spectrometer at the
Orph\'{e}e reactor. Three different experimental configurations were used:
the first is with a wavelength of 10 \AA\ and a sample-multidetector
distance of 5.6 m, allowing to study Q values between 4 10$^{-3}$ and 4 10$%
^{-2}$ \AA $^{-1}$ . The second one is with a wavelength of 4.5 \AA\ and a
sample detector distance of 5.6 m, allowing to reach higher values between 10%
$^{-2}$ and 10$^{-1}$ \AA $^{-1}$. The third one is with a wavelength of 4.5
\AA\ and a sample detector distance of 1.5 m, allowing to reach higher
values up to 4 10$^{-1}$ \AA $^{-1}$. The sample was introduced in a
cryostat with a superconducting split coil and aluminum windows. The
magnetic field is applied in an horizontal plane, perpendicular to the
neutron beam. Different orientations of the crystal were studied and we have
found that the scattering is roughly isotropic in this range of wave
vectors. Few parasitic reflections were removed before data treatment. In
order to subtract the background signal, an empty cell was measured. The
calibration of the spectrometer was performed with a Plexiglas sample
following the procedure given in reference\cite{neutronpa}. After
subtraction of the background, normalization by the Plexiglas sample, the
scattering function is presented in absolute units (cm$^{-1}$). We have
systematically neglected the inelastic spin wave corrections.

\section{Results}

\subsection{Magnetization versus magnetic field at 30K}

We have measured the magnetization of the sample as function of the magnetic
field in a Quantum Design PPMS magnetometer up to 5.9T (fig. 1). The
temperature of 30K was chosen because the ferromagnetic fraction of the
sample reaches 100\% at 5.9T. Let us describe this magnetic hysteresis
curve. Between 0 and 1T, the spins of the ferromagnetic part of the sample
are gradually oriented parallel to the applied field. Between 1 and 3T, the
sample is partly ferromagnet and partly antiferromagnet. The slope observed
in this range is mainly due to the susceptibility of the antiferromagnetic
part. Between 3 and 5.9T, the system transforms itself smoothly to pure
ferromagnetic system. When the magnetic field is removed, the system remains
ferromagnetic. Below 1T, ferromagnetic domains appear, leading to a zero
magnetization in zero field. From the fit of this curve, we have extracted
two important parameters, the ferromagnetic fraction $\phi $ and the
susceptibility of the antiferromagnetic part $\chi _{antif\text{ }}$ as
functions of the applied field B. Let us now describe the results of small
angle neutron scattering under the same experimental procedure.

\subsection{Orientation of the spins along the magnetic field}

As reported in the previous study\cite{nous}, the scattered intensity I(Q)
decreases in Q$^{-2}$ over a wide range of wave vectors Q, characteristic of
the behavior of a powder of 2D sheets without any correlations among them%
\cite{porod}. In this range of Q (0.02 to 0.1 \AA $^{-1}$), the small angle
scattering pattern at 30K in zero field (fig. 2 a) is isotropic. This is due
to the fact that the ferromagnetic domains and the domains of the phase
separation are both isotropic. The application of a magnetic field larger
than 2T ( fig. 2b at 2T) provides anisotropic iso-intensity curves which can
be fitted by parts of circles. This anisotropy is specific of the effect of
the orientation of the spins along the applied magnetic field. The intensity
I(Q,$\alpha $) is proportional to I(Q) sin$^2\alpha $ where Q is the modulus
of the scattering vector and $\alpha $ the angle between the vector Q and
the vector M (parallel to B)\cite{neutronpa}. In this range of Q, in which
the Q dependence is in Q$^{-2}$ (the ''cabbage structure''), the lines of
iso-intensity correspond to Q=Q$_0$sin$\alpha $, which are two parts of
circles. This is exactly what is observed here (as shown on fig 2b). This
very pecular behavior is thus explained. Consequently, one can say that the
application of a 2T magnetic field indeed corresponds to an orientation of
the spins along the magnetic field. There is no transformation of the phase
separation in this range of magnetic field, compatible with what was assumed
in the interpretation of the magnetization curve.

\subsection{Magnetic field induced evolution of the phase separation}

In order to analyze the Q dependence of the scattering function, we have
integrated the intensity over a $\mp $15 degrees cone for each value of Q.
In addition, we have assumed from the magnetization data that the compound
is completely ferromagnetic at 5.9T, so we have used the spectrum obtained
under 5.9T as a background for all the other results. The application of a
magnetic field, though it modifies strongly the orientation of the spins
(fig. 2), does not modify too strongly the shape of the curve (fig. 3a). We
have studied the magnetic field dependence of the scattering function in the
medium range of Q, where the Q$^{-2}$ dependence dominates. In this range of
Q, application of magnetic field decreases the scattered intensity. One can
notice that the curve at 2T is slightly different from the others (the slope
at small Q is smaller). We have no simple interpretation to this feature.

The same procedure was applied for the three different scattering
configurations that we used. One can note that the overlap between these
three configurations is very good. The results are shown on figure 3b. In
absence of magnetic field, the signal is very similar to that previously
published in the powdered sample, confirming that the scattering is mainly
isotropic. There is a nice Q$^{-2}$ dependence sample over a large range of
wave vectors. At small Q, the small angle scattering was dominated in the
powder by the granular structure: it obeys to the classical Porod law\cite
{porod} and varies in Q$^{-4}$. In the crystal, this contribution is much
smaller than in the powdered sample (as it should be) but the Q$^{-2}$
component remains the same. On this extended range of wavevectors, it is
clear that the slope at 4T is slightly different from that obtained in zero
field, suggesting that the shape of the domains is also slightly different.
This is not analyzed in the present work, but is related to a finite size of
the in-plane ferromagnetic domains (assumed to be infinite in the
''cabbage'' model), inducing a fractal dimension of the objects.

In a third part of the measurement, we have decreased the magnetic field and
found a signal which is completely different from the signal before and
during the application of the magnetic field (fig. 3c). At 2T, the signal
remains very small, indicating that the system remains mainly ferromagnetic.
At B=0, the Q$^{-2}$ component remains not visible, indicating that the
signal is not originating from ''cabbage'' phase separation but to classical
ferromagnetic quasi- isotropic Weiss domains (fig. 1).

The situation is summarized on fig. 1: Before the application of the
magnetic field, the phase separation presents the ''cabbage structure'' with
a stacking of domains ferro and antiferromagnetic. The ferromagnetic part
presents classical Weiss domains. Note that the model assumes that the
domains of the phase separation are infinite in two directions, but this is
only an approximation and it exits in the sample parts with all the possible
orientations of the sheets. Between 0 and 1T, the Weiss domains disappear.
Above 3T, the antiferromagnetic part is gradually transformed into
ferromagnetism. When the magnetic field is decreased, the system remains
ferromagnetic, but the Weiss domains nucleate back.

\subsection{Analysis of the absolute values: comparison to magnetization
measurements}

As discussed in a previous paper\cite{nous}, the Q$^{-2}$ dependence is that
of uncorrelated infinite 2D sheets\cite{porod}. From this point of view, the
''red cabbage'' structure can be slightly misleading. In the present case,
it corresponds to 2D ferromagnetic sheets in antiferromagnetic matrix which
plays the role of the vacuum. If one assumes that magnetic measurements
allow a precise determination of the ferromagnetic fraction $\phi ,$ the
only adjustable parameter in this model is the thickness of the 2D sheets
''t'' (total thickness of the ferro and antiferromagnetic layers). Then, the
thickness of the ferromagnetic sheets is $\phi $t and that of the
antiferromagnetic one is (1-$\phi $)t. For such a 2D object, the scattering
function is

I$_m$(Q)= 2$\pi $ $\phi (1-\phi )$t$_{eff}$ $\Delta \rho _m^2(B)$ Q$^{-2}%
\frac{(1-\cos Qt)}{(Qt)^2}$ sin$^2\alpha $ which is reduced in the case of a
wide dispersion of ''t'' values to:

I$_m$(Q)= $\pi \phi (1-\phi )$ t$_{eff}$ $\Delta \rho _m^2(B)$ Q$^{-2}$ sin$%
^2\alpha $

where t$_{eff}$ is given by t$_{eff}$ = t$\phi (1-\phi )$ and $\Delta \rho
_m(B)$ is the magnetic contrast which depends on B since the
antiferromagnetic scattering length is proportional to the applied field B. $%
\Delta \rho _m$ is proportional to the difference of magnetizations between
ferro and antiferromagnetic phases: $\Delta \rho _m$ =$%
x(M_{ferro}-M_{antif}).$ The amplitude of $M_{ferro}$ in this formula will
be assumed here to be constant 3.8 $\mu _{B\text{ }}$(its variation when the
magnetic field is applied is negligible) and $M_{antif}$ = $\chi _{antif%
\text{ }}$B is determined from magnetic measurements. The proportional
constant $x$ is 0.27 10$^{-12}$ cm/$\mu _B$ divided by the unit volume of a
formula unit 0.57 10$^{-22}$ cm$^3$, so $\Delta \rho _m^2$ =0.4 10$^{21}$ cm$%
^{-4}$ (1-$M_{antif}/M_{ferro}$)$^2$. Figure 4 presents $\phi $ and $\Delta
\rho _m(B)^2$ values extracted from fig. 1.

Experimentally, IQ$^2$=0.3 10$^{14}$ cm$^{-3}$in zero field. This drives,
using

I$_m$(Q)= $\pi \phi (1-\phi )$ t$_{eff}$ $\Delta \rho _m^2(B)$ Q$^{-2}$ sin$%
^2\alpha $

to a parameter value $\phi $t, which is about 15 \AA .

From these data, coupled to the $\phi $ and $\Delta \rho _m(B)$ values
extracted from fig. 1, it is possible to extract the parameter ''t'' and the
two parameters $\phi $t and (1-$\phi $)t, which are the thicknesses of
ferromagnetic and antiferromagnetic sheets respectively as functions of
magnetic field. They are shown on figure 5. On this figure, one can see that
the thickness of the ferromagnetic sheets increases versus magnetic field
and that the antiferromagnetic ones decreases, suggesting that only part of
the thickness of the antiferromagnetic sheets switches to ferromagnetic as
one increases the applied field. The simple picture in which the whole
thickness of the antiferromagnetic domain switches at once does not apply
here.

Recently, it was proposed \cite{fernandez} that the ''simple'' percolation
model cannot apply in the same compound since the ferromagnetic order
appears as long range order (larger than a few hundreds angtroms) in neutron
diffraction. However, this argument does not hold in the case of cabbage
structure as previously explained\cite{nous}. On the contrary, the present
results demonstrate that such a percolation model is really adequate.

\section{Conclusion}

In conclusion, the use of the magnetic SANS technique under magnetic field,
coupled to classical magnetization measurements allows to determine the
magnetic field dependence of the thickness of the different magnetic domains
in the phase separated Pr$_{1-x}$Ca$_x$MnO$_3$ with x=0.33 system. One can
see that this transformation corresponds to a gradual irreversible switching
from antiferromagnetic sheets to ferromagnetic ones, transforming the system
to complete ferromagnetic state at 6T. At about 3T, the system reaches the
percolation of the metallic ferromagnetic phase. This is at the origin of
the colossal magnetoresistance.

\section{Acknowledgments:}

We acknowledge L.\ Herv\'{e} for sample preparation, and very important
support from F.\ Ott and his cryostat. S. Mercone has been supported by a
Marie Curie fellowship of the European community program under contract
number HPMT2000-141. D.\ Saurel has been supported by the ''r\'{e}gion basse
Normandie''.

\section{Figure Captions}

Figure 1: The magnetization of the sample at 30K as function of magnetic
field (the field is ramped up to 5.9T and down to zero). Insets: schematic
drawings of the corresponding structures: before the application of the
field, the system is a coexistence at a nanoscopic scale of ferro (grey) and
antiferromagnetic (white) sheets. These sheets are supposed to be infinite
in the model, but different orientations of the sheets with respect to the
magnetic field (applied vertically in the pictures) are present in the
sample (''cabbage'' structure). The application of the magnetic field first
orientate the Weiss domains (up to 1T), then transforms the
antiferromagnetic phase into ferromagnetism (above 3T). At 5.9T, the system
is fully transformed. When the field is decreased, the system creates back
Weiss domains below 1T.

Figure 2: The scattering plane at 30K without and with a magnetic field of
2T. The magnetic field is applied horizontally in fig. b.

Figure 3 a: The magnetic small angle scattering function at 30K ramping up
from zero field to 5.9T for the intermediate range of wave vectors. The
signal obtained at 5.9T is used as background.

Figure 3 b: The small angle scattering function at 30K at 0T and 4T ramping
up from zero field. The three different experimental Q ranges are evidenced
by a small gap in the curves. On these curves, the background was also
chosen to be the sample itself at 5.9T where the whole sample is
ferromagnetic.

Figure 3 c: The small angle scattering function at 30K at 2T and zero field
after application of 6T, compared to the same curve before application of
field.

Figure 4 : $\phi $ and $\Delta \rho _m(B)^2$ values extracted from fig. 1.

Figure 5 : The magnetic field dependence at 30K of the two parameters t$%
_F=\phi $t and t$_{AF}=$(1-$\phi $)t, which are the thicknesses of the
ferromagnetic sheets and the antiferromagnetic sheets respectively.

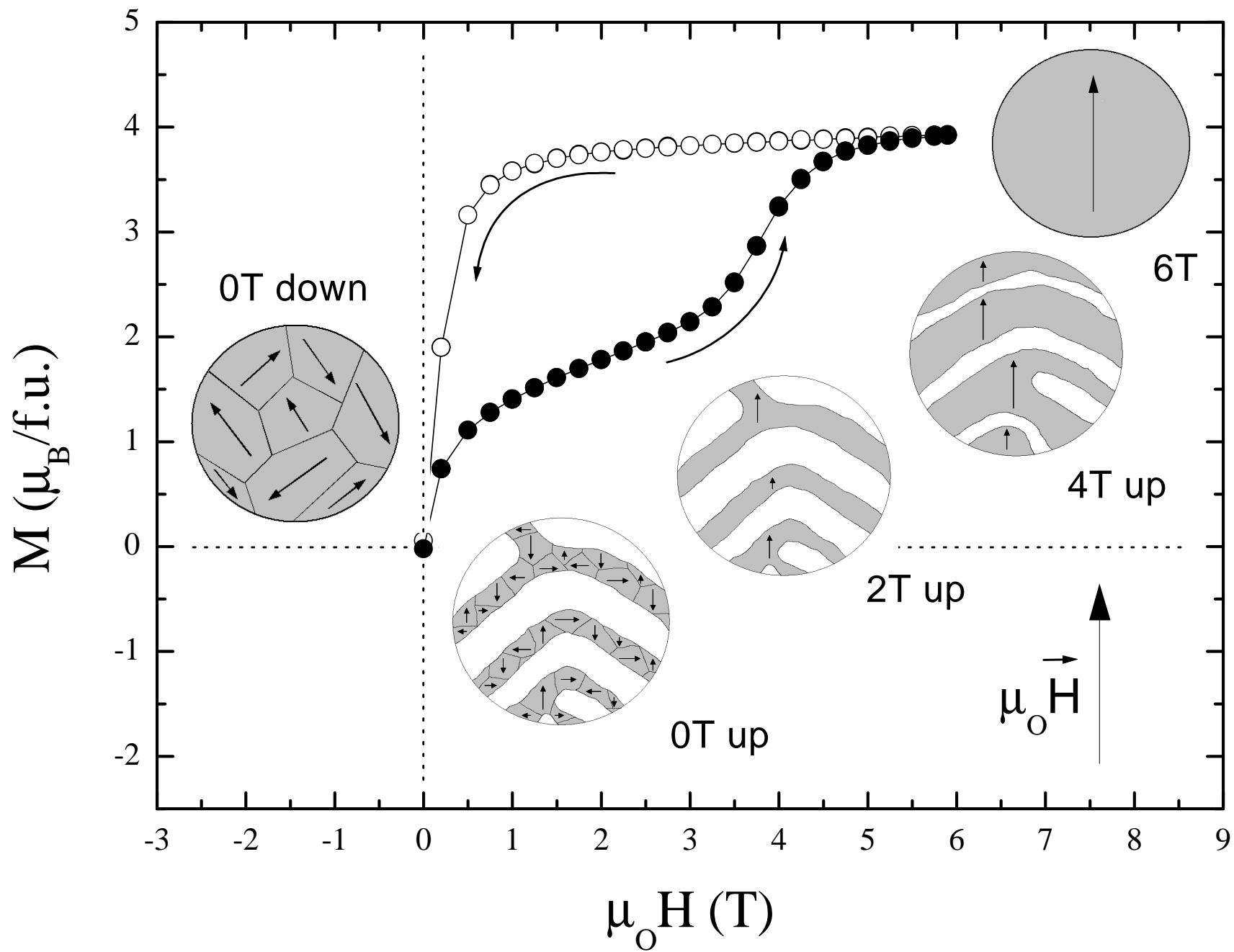
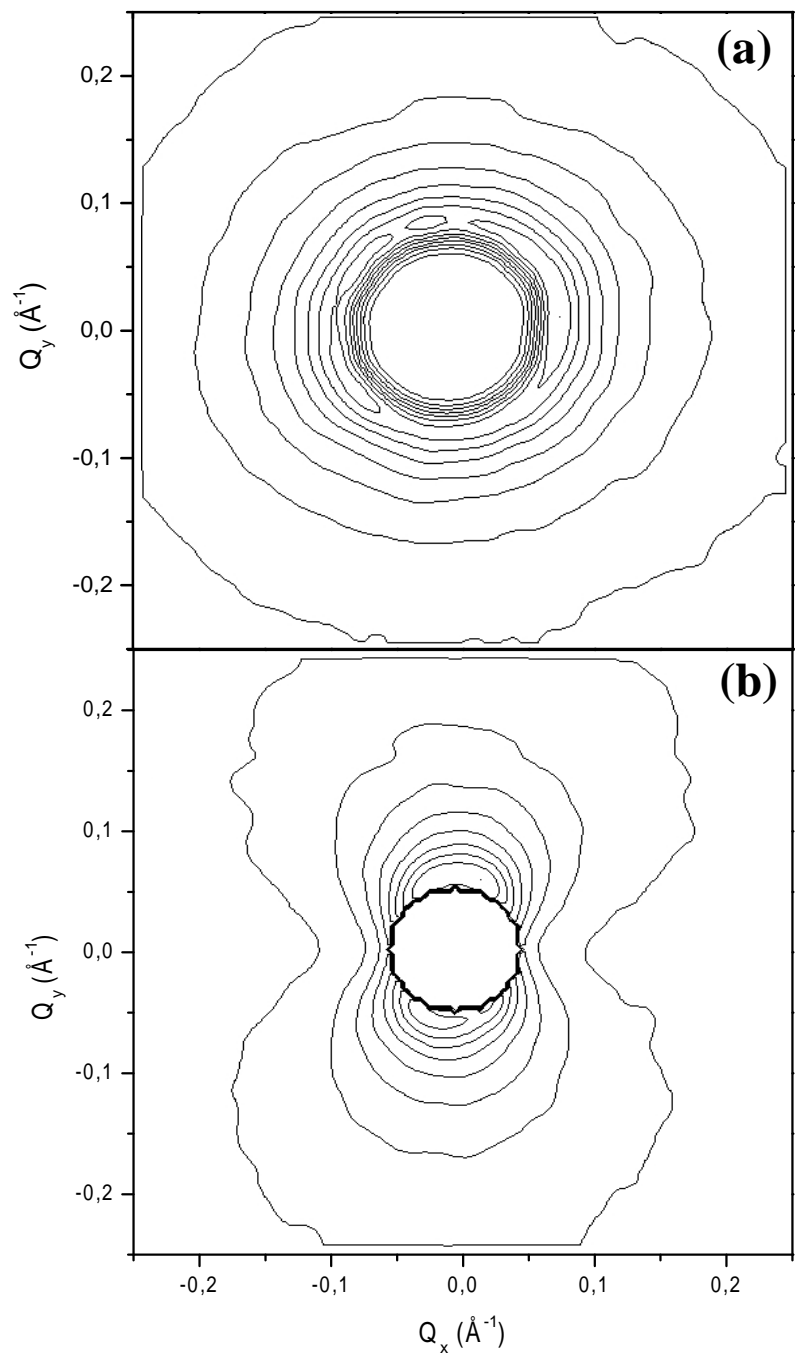
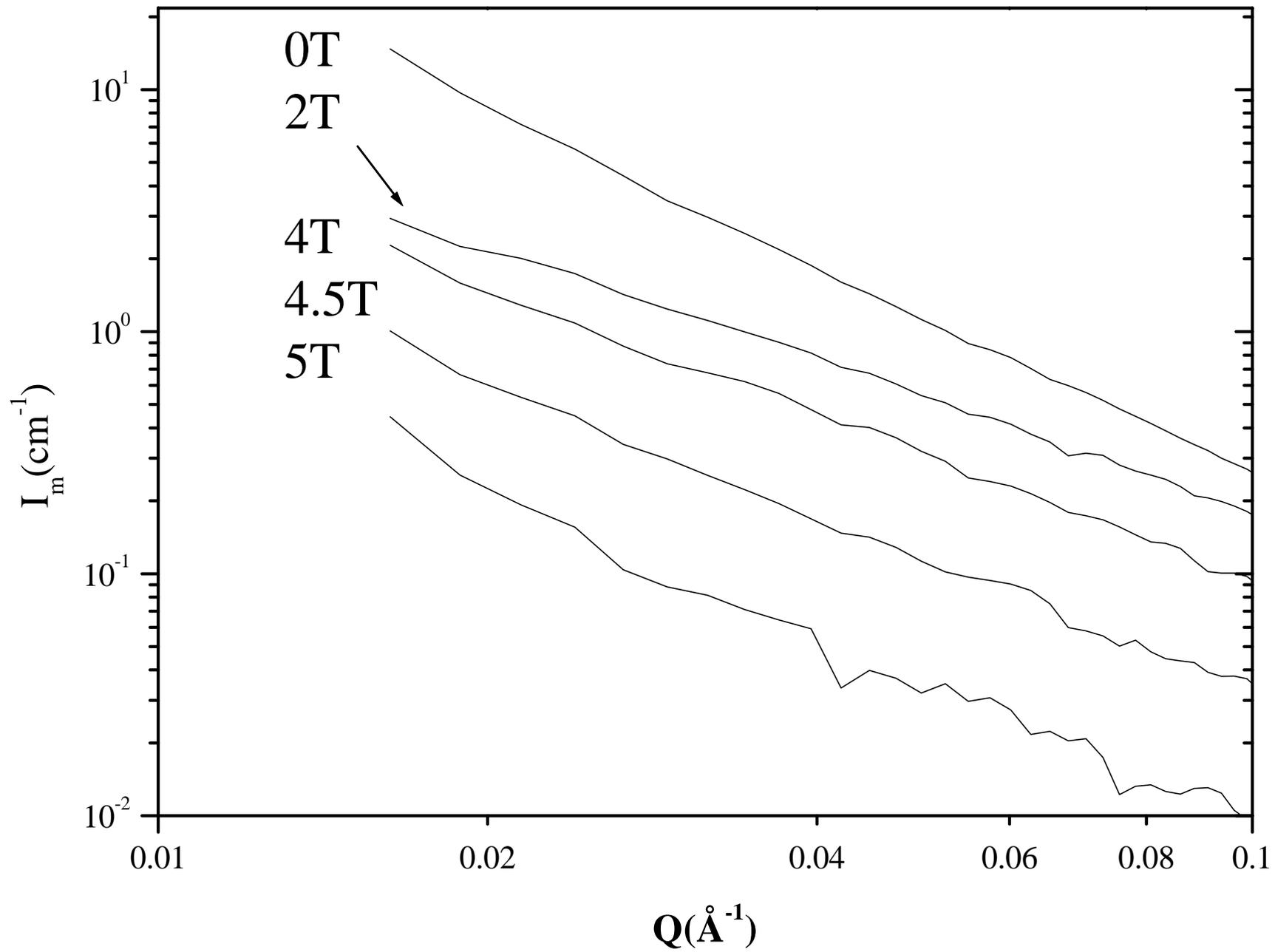
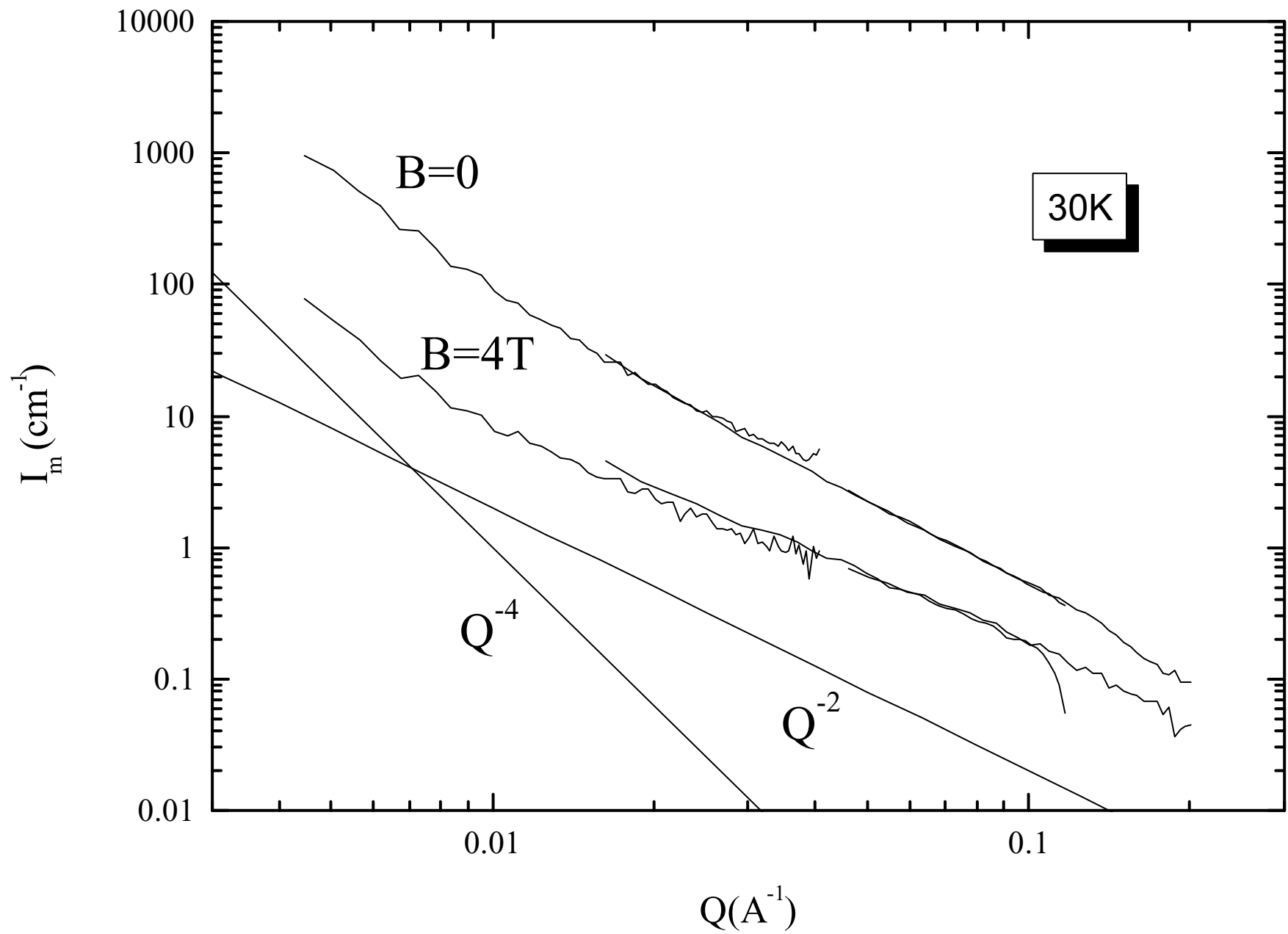
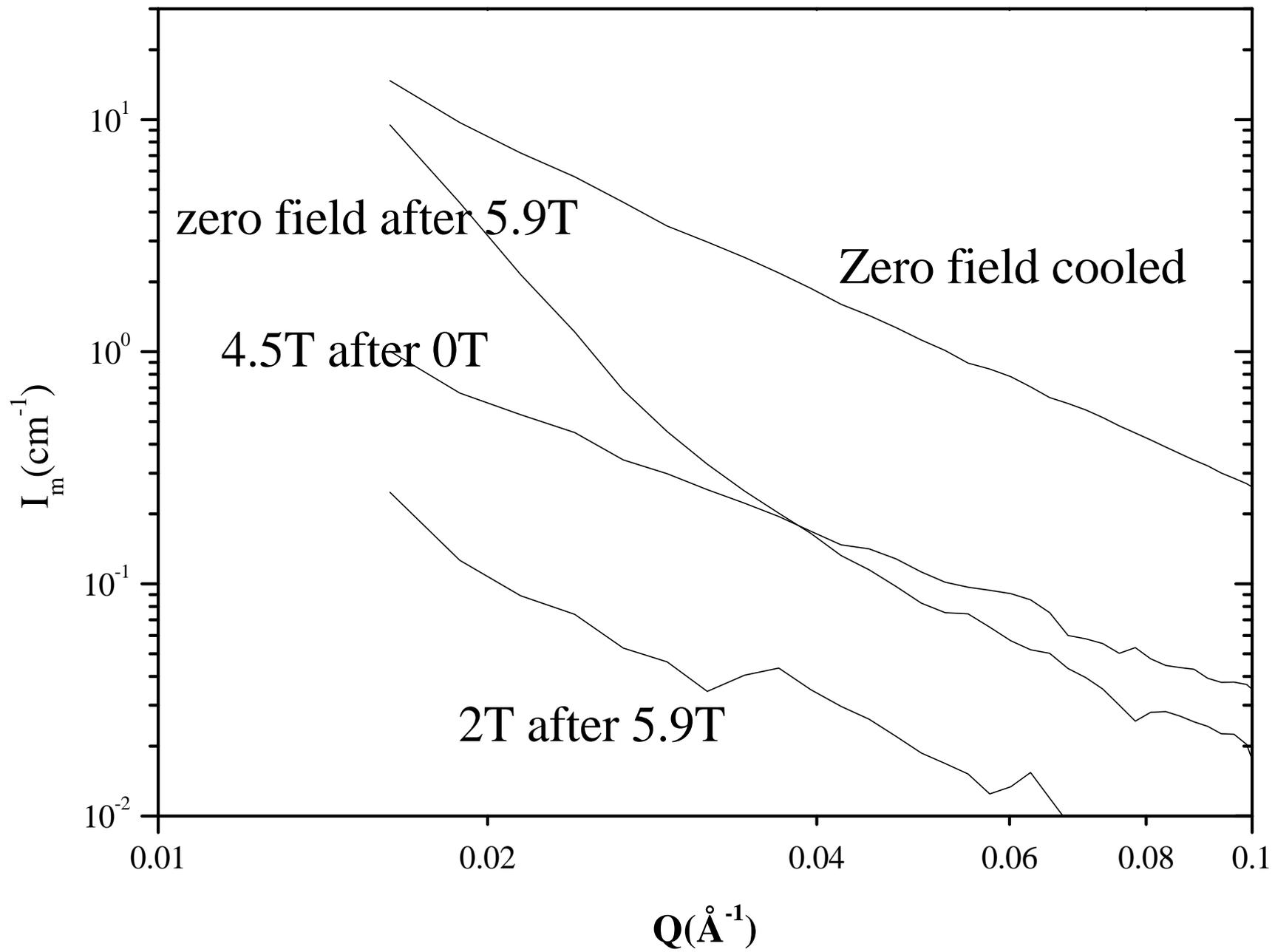
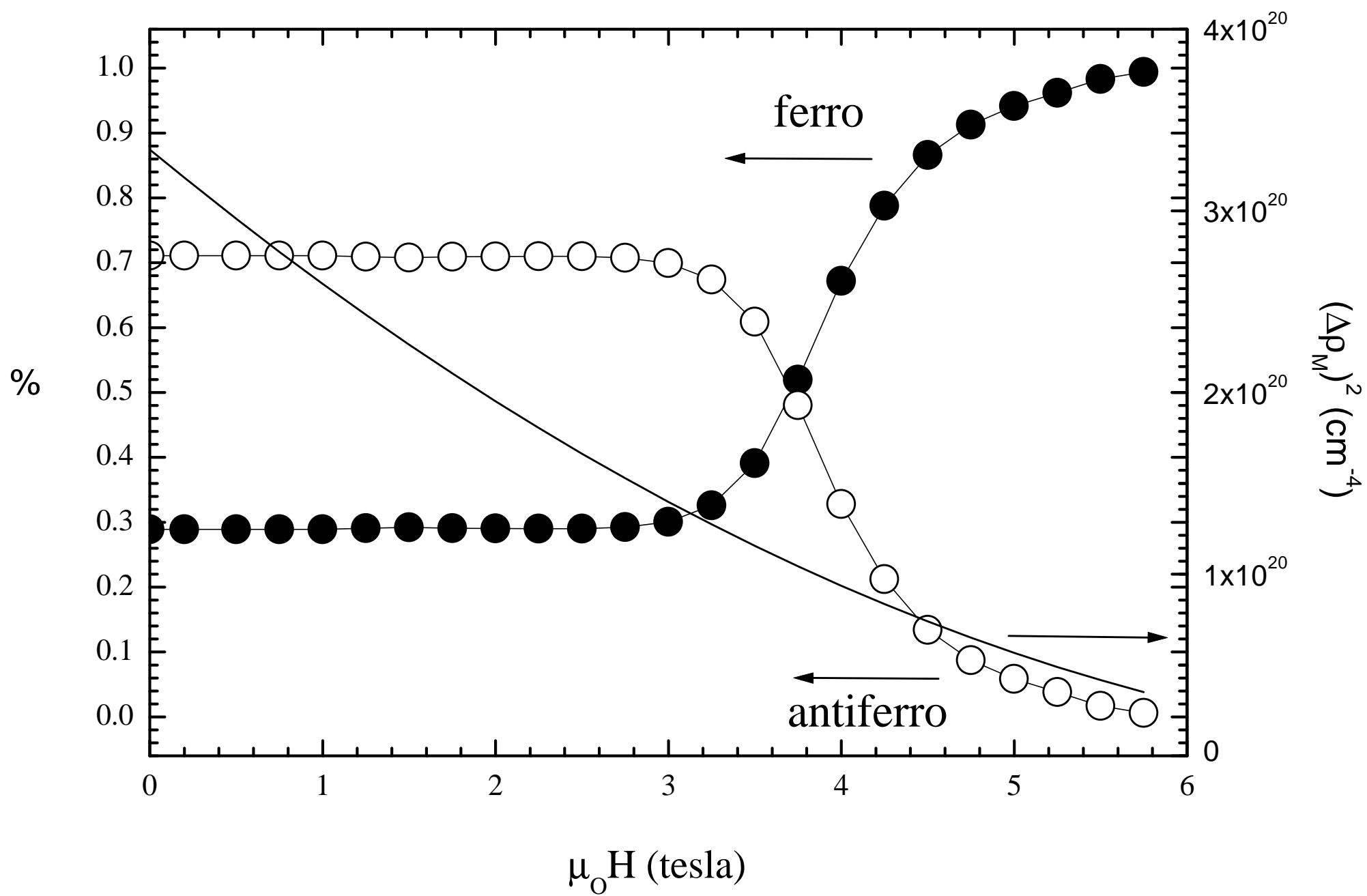
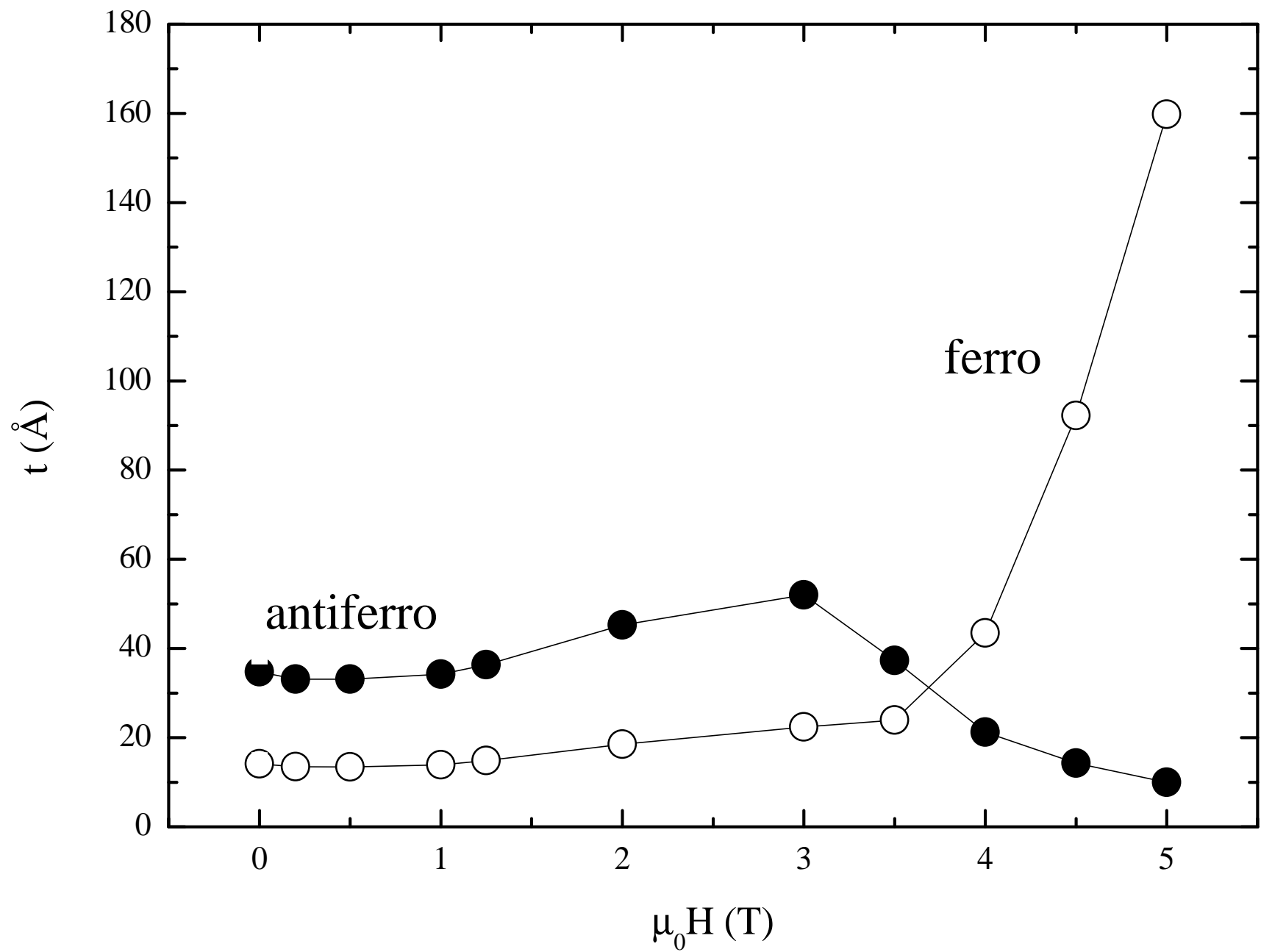
\end{document}